\documentclass[aps,prb,reprint,superscriptaddress,showpacs]{revtex4-1}

\usepackage{amssymb,amsfonts,latexsym,graphicx}
\usepackage{amsmath,bm}
\usepackage{hyperref}
\usepackage{booktabs}
\usepackage{mathtools}
\usepackage[ruled,vlined]{algorithm2e}
\usepackage{subfig}

\usepackage[normalem]{ulem} % For \xout, \sout Command

\newcommand{\E}{\ensuremath{\mathcal{E}}}
\newcommand{\e}{\ensuremath{\varepsilon}}
\newcommand{\kv}{{\boldsymbol k}}
\newcommand{\im}{\mathrm{i}} % For the imaginary unit

\begin{abstract}
  We demonstrate analytically and numerically that the dispersive Dirac
  cone emulating an epsilon-near-zero (ENZ) behavior is a universal
  property within a family of plasmonic crystals consisting of
  two-dimensional (2D) metals. Our starting point is a periodic array of 2D
  metallic sheets embedded in an {\em inhomogeneous and anisotropic}
  dielectric host that allows for propagation of transverse-magnetic (TM)
  polarized waves. By invoking a systematic bifurcation argument for
  arbitrary dielectric profiles in one spatial dimension, we show how TM
  Bloch waves experience an effective dielectric function that averages out
  microscopic details of the host medium. The corresponding effective
  dispersion relation reduces to a Dirac cone when the conductivity of the
  metallic sheet and the period of the array satisfy a critical condition
  for ENZ behavior. Our analytical findings are in excellent agreement with
  numerical simulations.
\end{abstract}

\begin{document}

\title{Universal behavior of dispersive Dirac cone in gradient-index
  plasmonic metamaterials}

\date{Draft of \today}

\author{Matthias Maier}
\thanks{\url{msmaier@umn.edu}; \url{http://www.math.umn.edu/~msmaier}}
\affiliation{School of Mathematics, University of Minnesota, Minneapolis, %
  Minnesota 55455, USA}

\author{Marios Mattheakis}
\thanks{\url{mariosmat@g.harvard.edu}; %
  \url{http://scholar.harvard.edu/marios_matthaiakis}}
\affiliation{School of Engineering and Applied Sciences, Harvard %
  University, Cambridge, Massachusetts 02138, USA}

\author{Efthimios Kaxiras}
\affiliation{School of Engineering and Applied Sciences, Harvard %
  University, Cambridge, Massachusetts 02138, USA}
\affiliation{Department of Physics, Harvard University, Cambridge, %
  Massachusetts 02138, USA}

\author{Mitchell Luskin}
\affiliation{School of Mathematics, University of Minnesota, Minneapolis, %
  Minnesota 55455, USA}

\author{Dionisios Margetis}
\affiliation{Department of Mathematics, and Institute for Physical Science
and Technology, and Center for Scientific Computation and Mathematical
Modeling, University of Maryland, College Park, Maryland 20742, USA.}

\maketitle

%%%%%%%%%%%%%%%%%%%%%%%%%%%%%%%%%%%%%%%%%%%%%%%%%%%%%%%%%%%%%%%%%%%%%%%%%%%%%%%%
%%%%%%%%%%%%%%%%%%%%%%%%%%%%%%%%%%%%%%%%%%%%%%%%%%%%%%%%%%%%%%%%%%%%%%%%%%%%%%%%

%%%%%%%%%%%%%%%%%%%%%%%%%%%%%%%%
%%%%%%%%%%%%%%%%%%%%%%%%%%%%%%%%
\section{Introduction}
\label{sec:Intro}
%%%%%%%%%%%%%%%%%%%%%%%%%%%%%%%%
%%%%%%%%%%%%%%%%%%%%%%%%%%%%%%%%

In the past few years the dream of manipulating the laws of optics at will
has evolved into a reality with the use of metamaterials. These structures
have made it possible to observe aberrant behavior like no refraction,
referred to as epsilon-near-zero (ENZ) \cite{prl97_2006, natMat10_2011,
natPhot7_2013, natPhot9_2015, mattheakis2016}, and negative refraction
\cite{wang2012}. This level of control of the path and dispersion of light
is of fundamental interest and can lead to exciting applications. In
particular, plasmonic metamaterials offer significant flexibility in tuning
permittivity or permeability values. This advance has opened the door to
novel devices and applications that include optical holography
\cite{nanoLet15_2015}, tunable metamaterials \cite{natNano10_2015,
prl116_2016}, optical cloaking \cite{science345_2014, pre72_2005}, and
subwavelength focusing lenses \cite{natNano6_2011,nanoLet14_2014}.

Plasmonic crystals, a class of particularly interesting metamaterials,
consist of stacked metallic layers arranged periodically with subwavelength
distance, and embedded in a dielectric host. These metamaterials offer new
`knobs' for controlling optical properties and can serve as
negative-refraction or ENZ media \cite{nature522_2015, prb90_2014,
prb91_2015}. The advent of truly two-dimensional (2D) materials with a wide
range of electronic and optical properties, comprising metals, semi-metals,
semiconductors, and dielectrics \cite{csr43_2014}, promise exceptional
quantum efficiency for light-matter interaction \cite{natPhot8_2014}. In
this paper, we characterize the ENZ behavior of a wide class of plasmonic
crystals by using a general theory based on Bloch waves.

The ultra-subwavelength propagating waves (plasmons) found in plasmonic
crystals based on 2D metals, in addition to providing extreme control over
optical properties \cite{prb87_2013, wang2014, prb80_2009, low17}, also
demonstrate low optical losses due to reduced dimensionality
\cite{mattheakis2016,wang2012}. In particular, graphene is a rather special
2D plasmonic material exhibiting ultra-subwavelength plasmons, and a high
density of free carriers which is controllable by chemical doping or bias
voltage \cite{prb80_2009, natPhot6_2012, nature487_2012, sharmila2017}. An
important finding is that the ENZ behavior introduced by subwavelength
plasmons is characterized by the presence of dispersive Dirac cones in
wavenumber space \cite{natMat10_2011, natPhot7_2013, natPhot9_2015,
mattheakis2016}. This linear iso-frequency dispersion relation was shown
for the special case of plasmonic crystals containing 2D dielectrics with
spatial-independent dielectric permittivity. This relation requires precise
tuning of system features~\cite{mattheakis2016}. It is not clear from this
earlier result to what extent the ENZ behavior depends on the homogeneity
of the 2D dielectric, or could be generalized to a wider class of
materials.

In this paper, we show that the occurrence of dispersive Dirac cones in
wavenumber space is a \emph{universal} property in plasmonic crystals with
dielectrics characterized by \emph{any spatial-dependent dielectric
permittivity} within a class of anisotropic materials. We provide an exact
expression for the critical structural period at which the multilayer
system behaves as an ENZ medium. This distance between adjacent sheets
depends on the permittivity profile of the dielectric host as well as on
the surface conductivity of the 2D metallic sheets. In addition, we give an
analytical derivation and provide computational evidence for our
predictions. To demonstrate the applicability of our approach, we
investigate numerically electromagnetic wave propagation in \emph{finite}
multilayer plasmonic structures, and verify the ENZ behavior at the
predicted structural period. These results suggest a systematic approach to
making general and accurate predictions about the optical response of
metamaterials based on 2D multilayered systems. An implication of our
method is the emergence of an \emph{effective} dielectric function in the
dispersion relation, which can be interpreted as the result of an averaging
procedure (homogenization). This view further supports the universal
character of our theory.

The remainder of the paper is organized as follows. In
Sec.~\ref{sec:Bloch-wave}, we introduce the problem geometry and general
formulation by Bloch-wave theory. Section~\ref{sec:parabolic} outlines the
exactly solvable example of parabolic permittivity of the dielectric host.
In Sec.~\ref{sec:universal}, we develop a bifurcation argument that
indicates the universality of the dispersion relation and ENZ behavior for
a class of plasmonic crystals. Section~\ref{sec:discussion} concludes our
analysis by pointing out a linkage of our results to the homogenization of
Maxwell's equations. The $e^{-\im\omega t}$ time dependence is assumed
throughout, where $\omega$ is the angular frequency. We write $f=\mathcal
O(h)$ to imply that $|f/h|$ is bounded in a prescribed limit.

%%%%%%%%%%%%%%%%%%%%%%%%%%%%%%%%%%%%%%%%%%%%%%%%%%%%%%%%%%%%%%%%%%%%%%%%%%%%%%%%
%%%%%%%%%%%%%%%%%%%%%%%%%%%%%%%%%%%%%%%%%%%%%%%%%%%%%%%%%%%%%%%%%%%%%%%%%%%%%%%%
%%%%%%%%%%%%%%%%%%%%%%%%%%%%%%%%%%%%%%%%%%%%%
%%%%%%%%%%%%%%%%%%%%%%%%%%%%%%%%%%%%%%%%%%%%%
\section{Geometry and Bloch-wave theory}
\label{sec:Bloch-wave}
%%%%%%%%%%%%%%%%%%%%%%%%%%%%%%%%%%%%%%%%%%%%%
%%%%%%%%%%%%%%%%%%%%%%%%%%%%%%%%%%%%%%%%%%%%%

\begin{figure}
  \centering
  \includegraphics[scale=1.0]{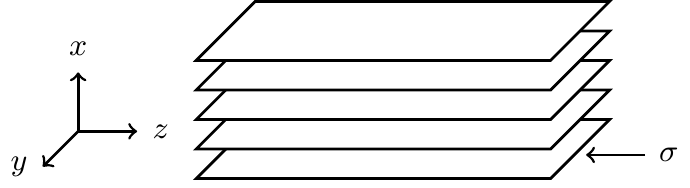}
  \caption{\label{fig:layered-structure}
    Geometry of the plasmonic crystal. The layered structure is periodic in
    the $x$-direction and consists of planar 2D metallic sheets with
    isotropic conductivity $\sigma$.}
\end{figure}

In this section, we describe the geometry of the problem and the related
Bloch-wave theory. Consider a plasmonic crystal that is periodic in the
$x$-direction and consists of flat 2D metallic sheets with isotropic
surface conductivity $\sigma$ (see Fig.~\ref{fig:layered-structure}). Each
sheet is parallel to the $yz$-plane and positioned at $x=nd$ for integer
$n$.

The material filling the space between any two consecutive sheets is
described by the anisotropic relative permittivity tensor
$\text{diag}\,(\e_x,\e_y,\e_z)$, where $\e_x=\text{constant}$, and $\e_y(x)
= \e_z(x)$ depends on the spatial coordinate $x$ with period $d$. Here, we
set the vaccuum permittivity equal to unity, $\e_0=1$. We seek solutions of
time-harmonic Maxwell's equations with transverse-magnetic (TM)
polarization, that is, with electric and magnetic field components
${\boldsymbol E}=(E_x,0,E_z)$ and ${\boldsymbol H}=(0, H_y, 0)$. The
assumed TM-polarization and the symmetry of the physical system suggest
that
\begin{align*}
  E_z(x,z) = \mathcal{E}(x)\,e^{\im k_zz},
\end{align*}
which effectively reduces the system of governing equations to a 2D
problem. Substituting the above ansatz into time-harmonic Maxwell's
equations and eliminating $E_x$ and $H_y$ leads to the following ordinary
differential equation for $\E(x)$:
\begin{align}
  -\partial_x^2\E + \kappa(k_z)\e_z(x) \E = 0,
  \quad
  \kappa(k_z)=\frac{k_z^2-k_0^2\e_x}{\e_x},
  \label{eq:governing_ode}
\end{align}
where $\mu$ denotes the permeability of the ambient material and
$k_0=\omega\sqrt{\mu}$. By the continuity of the tangential electric field
and the jump discontinuity of the tangential magnetic field due to surface
current, the metallic sheets give rise to the following transmission
conditions at $x=nd$:
\begin{align*}
  \begin{cases}
    \E^+=\E^-,
    \\[0.3em]
   -\im(\omega/\sigma)
    \left[\big(\partial_x\E\big)^+-\big(\partial_x\E\big)^-\right]
    = \kappa(k_z)\,\E^+,
  \end{cases}
\end{align*}
where $(.)^{\pm}$ indicates the limit from the right ($+$) or the left
($-$) of the metallic boundary. In order to close the system of equations,
we make a Bloch-wave ansatz in the $x$-direction, with $k_x$ denoting the
real Bloch wavenumber:
\begin{align*}
  \E(x) = e^{\im k_x d}\E(x-d),
  \quad
  \partial_x\E(x) = e^{\im k_x d}\partial_x\E(x-d).
\end{align*}
The combination of the transmission conditions and the periodicity
assumption leads to a closed system consisting of
Eq.~\eqref{eq:governing_ode} and the following boundary conditions:
\begin{align*}
  \begin{bmatrix}\E(d^-) \\ \E'(d^-)\end{bmatrix}
  =
  e^{\im k_x d}
  \begin{bmatrix}
    1 & 0 \\
    - \im (\sigma/\omega)\kappa(k_z)& 1
  \end{bmatrix}
  \begin{bmatrix}\E(0^+) \\ \E'(0^+)\end{bmatrix}\;,
\end{align*}
with $\E'(x)=\partial_x\E(x)$.

We next describe the \emph{dispersion relation} between $k_x$ and $k_z$ in
general terms. In the following analysis, we work in the 2D wavenumber
space with $\kv=(k_x,k_z)$. To render Eqs.~\eqref{eq:governing_ode} with
the above boundary conditions amenable to analytical and numerical
investigation, we perform an additional algebraic manipulation: Let
$\E_{(1)}(x)$ and $\E_{(2)}(x)$ be solutions of
Eq.~\eqref{eq:governing_ode} with initial conditions
\begin{align}
  \E_{(1)}(0) = 1, \; \E'_{(1)}(0) = 0,
  \quad
  \E_{(2)}(0) = 0, \; \E'_{(2)}(0) = 1.
  \label{eq:initial_conditions}
\end{align}
These solutions are linearly independent and therefore the general solution
for $\E(x)$ is given by $\E(x)=c_1\E_{(1)}(x)+c_2\E_{(2)}(x)$. The
existence of a non-trivial solution implies the condition
\begin{multline}
  D[\kv] = \det
  \Bigg(
    e^{\im k_x d}
    \begin{bmatrix}
      1 & 0 \\
      - \im (\sigma/\omega)\kappa(k_z)& 1
    \end{bmatrix}
    \\
    -
    \begin{bmatrix}
      \E_{(1)}(d) & \E_{(2)}(d) \\
      \E'_{(1)}(d) & \E'_{(2)}(d)
    \end{bmatrix}
  \Bigg)
  \;=\; 0.
  \label{eq:determinant_condition}
\end{multline}
Equation~\eqref{eq:determinant_condition} expresses an implicit dispersion
relation, namely, the locus of points $\kv$ such that $D[\kv]=0$.

%%%%%%%%%%%%%%%%%%%%%%%%%%%%%%%%%%%%%%%%%%%%%%%%%%%%%%%%%%%%%%%%%%%%%%%%%%%%%%%%
%%%%%%%%%%%%%%%%%%%%%%%%%%%%%%%%%%%%%%%%%%%%%%%%%%%%%%%%%%%%%%%%%%%%%%%%%%%%%%%%
%%%%%%%%%%%%%%%%%%%%%%%%%%%%%%%%%%%%%%%%%%%%%%%%%%%%%%%%%%%%%%%%%%
%%%%%%%%%%%%%%%%%%%%%%%%%%%%%%%%%%%%%%%%%%%%%%%%%%%%%%%%%%%%%%%%%%
\section{An example: Parabolic dielectric profile}
\label{sec:parabolic}
%%%%%%%%%%%%%%%%%%%%%%%%%%%%%%%%%%%%%%%%%%%%%%%%%%%%%%%%%%%%%%%%%%
%%%%%%%%%%%%%%%%%%%%%%%%%%%%%%%%%%%%%%%%%%%%%%%%%%%%%%%%%%%%%%%%%%

For certain permittivity profiles $\e_z(x)$ of period $d$, the system of
Eqs.~\eqref{eq:governing_ode} and \eqref{eq:initial_conditions} admits
exact, closed-form solutions. Thus, Eq.~\eqref{eq:determinant_condition}
is made explicit. Next, we present analytical and computational results for
a parabolic permittivity profile $\e_z(x)$. Note that the case of constant
permittivity, $\e_z(x)=\,\text{const.}$, is analyzed in
Ref.~\onlinecite{mattheakis2016}.

\begin{figure*}
  \begin{center}
      \mbox{
      \subfloat[]{\includegraphics[scale=0.09]{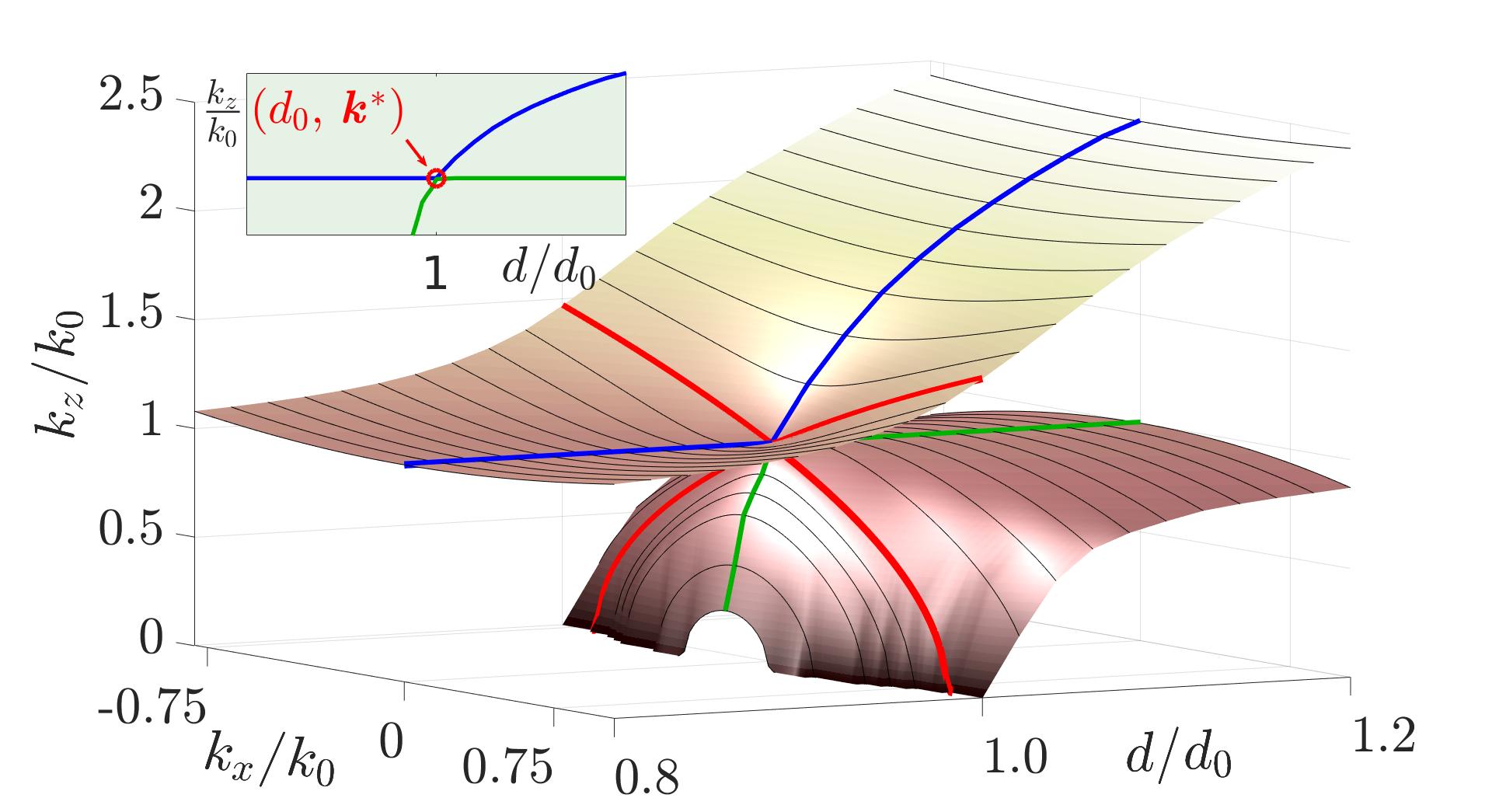}}
      \hspace{-1.1em}
      \subfloat[]{\includegraphics[scale=0.09]{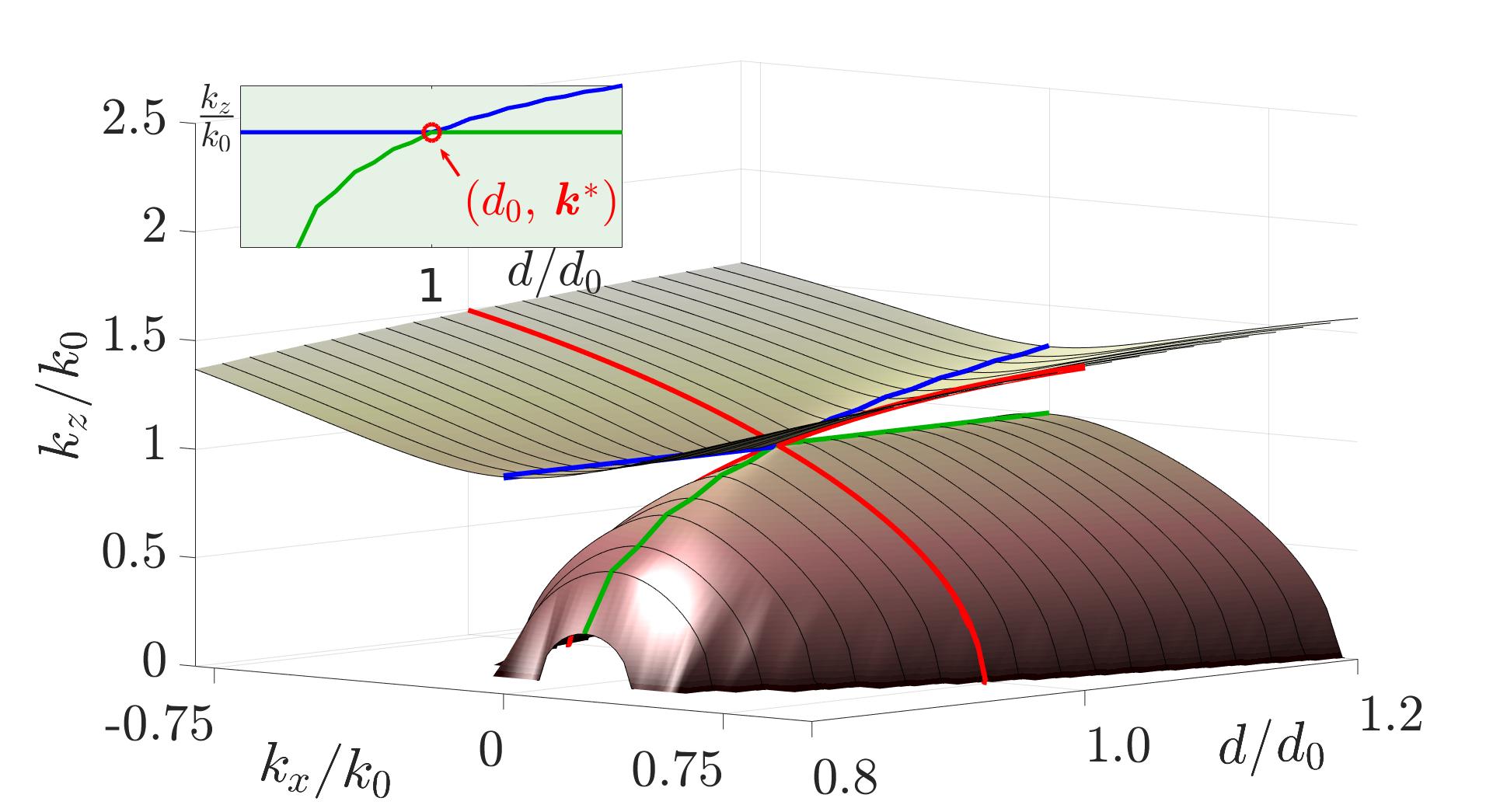}}
      \hspace{-1.1em}
      \subfloat[]{\includegraphics[scale=0.09]{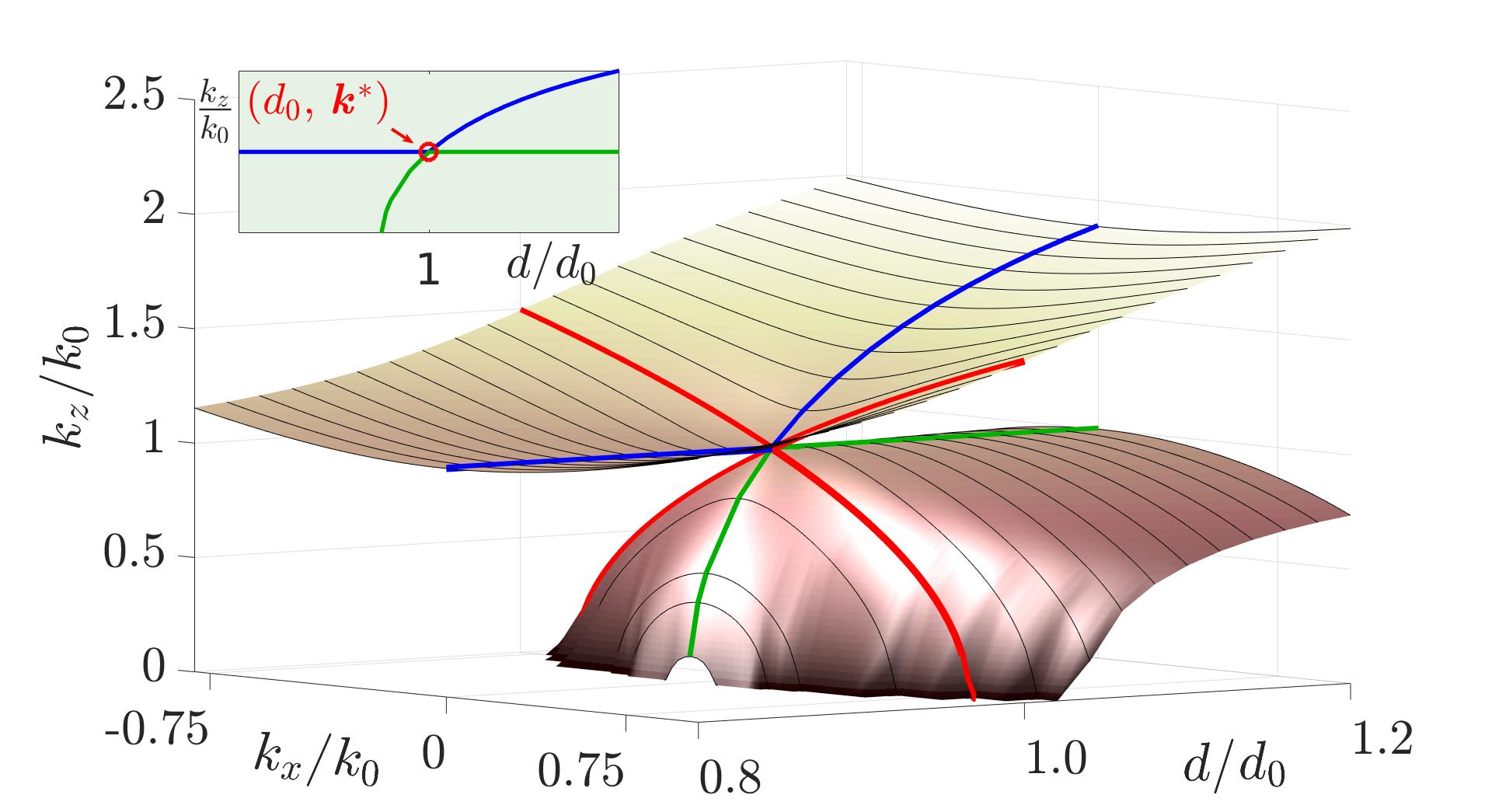}}}
  \end{center}
  \caption{\label{fig:parameter_study} Parameter study of numerically
    computed dispersion curves for (a) the parabolic profile
    \eqref{eq:parabolic_profile} with the scaling parameter $\alpha=20/3$,
    (b) double-well profile $f_{\text{dw}}$ and (c) non-symmetric profile
    $f_{\text{ns}}$. $d/d_0$ is chosen in the range from 0.8 to 1.2. The
    red solid lines indicate the Dirac cones dispersion at $d=d_0$. The
    inset shows the dispersion relation at the center of the Brillouin zone
  $\kv^\ast$ as function of $d$ (blue and green cutlines in the major
image).}
\end{figure*}
Accordingly, consider the \emph{parabolic} dielectric profile
\begin{align}
  \e_z(x)=\e_{z,0}\Big[1+6\,\alpha\,\frac xd\,\Big(1-\frac xd\Big)\Big],
  \label{eq:parabolic_profile}
\end{align}
which is well known in optics \cite{parabolicIndexBook, marios2012}. Here,
$\alpha>0$ is a scaling parameter with background dielectric permittivity
$\e_{z,0}>0$. In this case, $\mathcal{E}_{(1)}$ and $\mathcal{E}_{(2)}$ can
be written in terms of closed-form special functions (see
Appendix~\ref{app:parabolic}). Relation \eqref{eq:determinant_condition}
can be further simplified in the vicinity of the center of the Brillouin
zone, where $|k_x d|\ll 1$, by choosing the branch of the dispersion
relation containing $\kv^\ast=(k_x^\ast,k_z^\ast)=(0,\pm k_0\sqrt{\e_x})$.
As a result of this simplification, the Bloch wave sees a homogeneous
medium with effective permittivity
$\e=\text{diag}\,\big(\e_x,\e_z^{\text{eff}},\e_z^{\text{eff}}\big)$. The
dispersion relation is
\begin{align}
  \frac{k_x^2}{\e_z^{\text{eff}}}
  \;+\;
  \frac{k_z^2}{\e_{x}}
  \;=\;
  k_0^2,
  \qquad
  \frac{\e_z^{\text{eff}}}{\e_{z,0}}
  =
  1+\alpha - \frac{\xi_0}{d}.
  \label{eq:effective_dispersion_relation}
\end{align}
In the above, $\xi_0$ denotes the plasmonic thickness, $\xi_0=-\im
\sigma/(\omega\e_{z,0})$ \cite{mattheakis2016, wang2012, wang2014}. Here,
we assume for the sake of argument that $\sigma$ is a purely imaginary
number so that $\xi_0$ is real valued. Below, we will provide a derivation
of \eqref{eq:effective_dispersion_relation} for general profiles $\e_z(x)$.
Dispersion relation \eqref{eq:effective_dispersion_relation} is valid in a
neighborhood of $\kv^\ast$. For $\e_z^{\text{eff}}\gtrless0$, this relation
describes an elliptic, or hyperbolic band, respectively.

The ENZ behavior is characterized by $\e_z^{\text{eff}}\approx0$ in
dispersion relation \eqref{eq:effective_dispersion_relation}
\cite{mattheakis2016}. In the case of the parabolic profile of this
section, this condition is achieved if $\xi_0/d = 1+\alpha$. This motivates
the definition of the critical ENZ structural period,
\begin{align}
  d_0 = \xi_0/\big(1+\alpha\big).
  \label{eq:eps_near_zero_condition}
\end{align}
A breakdown of Eq.~\eqref{eq:effective_dispersion_relation} due to
$\e_z^{\text{eff}}=0$ is a necessary condition to observe linear dispersion
and thus dispersive Dirac cones~\cite{mattheakis2016}. Even though
Eq.~\eqref{eq:effective_dispersion_relation} is an approximate formula
describing the dispersion relation in the neighborhood of $\kv^\ast$, the
ENZ condition $d=d_0$ is \emph{exact} for the existence of a Dirac
cone for this example of a parabolic profile.

In the case with a lossy metallic sheet, when $\sigma$ has positive real
part, $d_0$ becomes a complex-valued number an, thus, the ENZ condition
$d=d_0$ cannot be satisfied exactly. However, for all practical purposes,
losses are typically very small such that an effective ENZ behavior can be
approximately observed with the choice $d=\textrm{Re}(d_0)$.

We now verify the effective theory given by
Eqs.~\eqref{eq:effective_dispersion_relation} and
\eqref{eq:eps_near_zero_condition} numerically. In order to compute all
real-valued dispersion bands located near $\kv^\ast$, we solve the system
of Eqs.~\eqref{eq:governing_ode}, \eqref{eq:initial_conditions}, and
\eqref{eq:determinant_condition} (for details see
Appendix~\ref{app:Newton_scheme}). We carry out a parameter study with  the
scaling parameter $\alpha=20/3$, background  permittivity components
$\e_{z,0}=2$ (in-plane) and $\e_x=1$ (out-of-plane), and $d/d_0$ in the
range from 0.8 to 1.2. The numerically computed dispersion bands are shown
in Fig.~\ref{fig:parameter_study}a. A band gap appears for values of $d$
different than $d_0$.

%%%%%%%%%%%%%%%%%%%%%%%%%%%%%%%%%%%%%%%%%%%%%%%%%%%%%%%%%%%%%%%%%%%%%%%%%%%%%%%%
%%%%%%%%%%%%%%%%%%%%%%%%%%%%%%%%%%%%%%%%%%%%%%%%%%%%%%%%%%%%%%%%%%%%%%%%%%%%%%%%
%%%%%%%%%%%%%%%%%%%%%%%%%%%%%%%%%%%%%%%%%%%%%%%%%%%%%%%%%%%%%%%%%%
%%%%%%%%%%%%%%%%%%%%%%%%%%%%%%%%%%%%%%%%%%%%%%%%%%%%%%%%%%%%%%%%%%
\section{Universality of dispersion relation and ENZ condition}
\label{sec:universal}
%%%%%%%%%%%%%%%%%%%%%%%%%%%%%%%%%%%%%%%%%%%%%%%%%%%%%%%%%%%%%%%%%%
%%%%%%%%%%%%%%%%%%%%%%%%%%%%%%%%%%%%%%%%%%%%%%%%%%%%%%%%%%%%%%%%%%
In this section, we address the problem of arbitrary $\e_z(x)$, both
analytically and numerically. We claim that effective dispersion relation
\eqref{eq:effective_dispersion_relation} and ENZ condition
\eqref{eq:eps_near_zero_condition} are in fact \emph{universal} within the
model of Sec.~\ref{sec:Bloch-wave}. This means that they are valid for any
tensor permittivity $\text{diag}\,(\e_x,\e_z,\e_z)$ with arbitrary,
spatial-dependent $\e_z(x)$. To develop a general argument, we set
\begin{align}
  \e_z(x)=\e_{z,0}f(x/d),
  \quad
  f(x)>0,
  \label{eq:general_profile}
\end{align}
where $f(x)$ is an arbitrarily chosen, continuous and periodic positive
function. Guided by our results for the parabolic profile
(Sec.~\ref{sec:parabolic}), we now make the conjecture that dispersion
relation \eqref{eq:effective_dispersion_relation} still holds with the
definitions
\begin{align}
  d_0 = \xi_0 \left [\int_0^1f(x)\text{d}x\right]^{-1},
  \quad
  \frac{\e_z^{\text{eff}}}{\e_{z,0}}=
  \xi_0 \left( \frac{1}{d_0}
  - \frac{1}{d}\right).
  \label{eq:general_eeff}
\end{align}

In the following analysis, we give a formal bifurcation argument
justifying definition \eqref{eq:general_eeff}. We start by expanding
Eq.~\eqref{eq:determinant_condition} in the neighborhood of $\kv^\ast$ in
powers of the components of $\delta \kv =(\delta k_x,\delta k_z) = \kv^\ast
- \kv$. First, it can be readily shown that at $\kv=\kv^\ast$
Eq.~\eqref{eq:governing_ode} reduces to $-\partial_x^2\E=0$. Thus, the
system of fundamental solutions is given by $\E_{(1)}(x)=1$,
$\E_{(2)}(x)=x$. This implies that $D[\kv^\ast]=0$. The expansion of
$D[\kv]$ up to second order in $\delta \kv$ leads to an expression of the
form
\begin{multline*}
  D[\kv^\ast+\delta\kv] =
  \\
  b_x\delta k_x + b_z\delta k_z
  + b_{xx} (\delta k_x)^2 + b_{zz} (\delta k_z)^2 + b_{xz} \delta k_x\delta
  k_z.
\end{multline*}
\emph{The occurrence of a Dirac point is identified with the appearance of
a critical point for $D[\kv]$, when $b_x=b_z=0$.} In order to express $b_x$
and $b_z$ in terms of physical parameters, we notice that only the term
$\big[
\im e^{\im k_xd}(\sigma/\omega)\kappa(k_z)+
\E'_{(1)}(d)\big]\,\E_{(2)}(d)$ of $D[(k_x,k_z)]$ contributes to
first order in $\delta\kv$. Accordingly, we find
\begin{multline*}
  D[\kv^\ast+\delta\kv] =
  \\
  -d\,\Big(\frac{-\im \sigma}{\omega\e_x}\,2k_z\delta k_z -
  \delta\E'_{(1)}(d)[\delta k_z]\Big)+\mathcal{O}((\delta \kv)^2).
\end{multline*}
Here, $\delta\E[\delta k_z]$ denotes the total variation of $\E$ with
respect to $k_z$ in the direction $\delta k_z$. It can be shown (see
Appendix~\ref{app:Newton_scheme}) that $\delta\E_{(1)}[\delta k_z]$ solves the
differential equation
$-\partial^2_x\delta\E_{(1)}=-\e_z(x)/\e_x\,2k_z\delta k_z$.
The solution has the derivative
\begin{align*}
  \delta\E'_{(1)}(x)=2k_z\delta k_z \left [
  \e_x\,\int_0^x\e_z(\xi)\,\text{d}\xi \right ]^{-1},
\end{align*}
which enters $D[\kv^\ast+\delta\kv]$. Thus, we obtain $b_x=0$
and
\begin{equation*}
  b_z
  =\left [ \xi_0-d\int_0^1 f(x)\text{d}x\right ]
  \frac{2dk_z\e_{z,0}}{\e_x}.
\end{equation*}
At the critical point, the expression in the bracket must vanish,
which produces Eq.~\eqref{eq:general_eeff}.

A refined computation for the critical case of $d=d_0$ gives
$b_{xx}=-d^2$, $b_{xz}=0$, and $b_{zz}>0$. Thus, the effective
dispersion relation at $d/d_0=1$ up to second-order terms is
$b_{xx}\delta k_x^2+b_{zz}\delta k_z^2=0$ with $b_{xx}b_{zz}<0$, which
corresponds to a Dirac cone. Moreover, for $\e_z^{\text{eff}}/\e_x\sim1$ it
can be shown that
\begin{equation*}
  D[\kv^\ast+\delta\kv] \approx
  -d^2\Big[\delta k_x^2 + \frac{\e_z^{\text{eff}}}{\e_x}(k_z^\ast+\delta
  k_z)^2 - \frac{\e_z^{\text{eff}}}{\e_x}(k_z^\ast)^2\Big].
\end{equation*}
By $(k_z^\ast)^2/\e_x = k_0^2$, the above relation recovers the
elliptic profile of Eq.~\eqref{eq:effective_dispersion_relation}.

\begin{figure}[tbph!]
  \centering
  \includegraphics[scale=1.0]{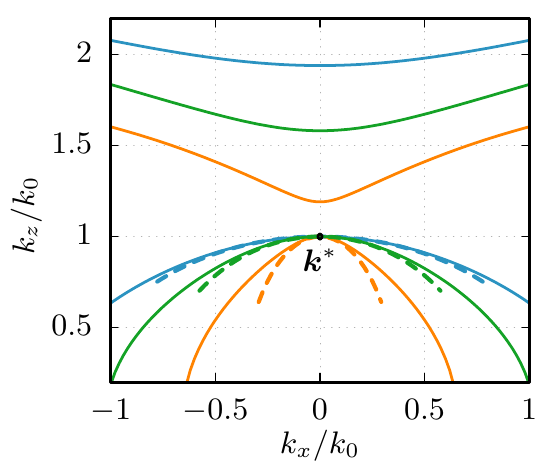}
  \caption{\label{fig:enz_and_approx}
    Numerically computed dispersion curves (solid lines) and the effective
    dispersion relation Eq.~\ref{eq:effective_dispersion_relation} (dashed
    lines) for $d/d_0=1.1$ computed for the parabolic (blue), double-well
    $f_{\text{dw}}(x)$ (orange), and nonsymmetric profile
    $f_{\text{ns}}(x)$ (green). The curvatures agree at the critical point
  $\kv^\ast$.}
\end{figure}
In order to support this bifurcation argument with numerical evidence, we
test Eq.~\eqref{eq:general_eeff} for two additional dielectric profiles
which to our knowledge do not admit exact solutions in simple closed form.
In the spirit of Ref.~\onlinecite{roger78}, we study distinctly different
profiles $\e_z(x)$. Specifically, we use the symmetric double-well profile
$f_{\text{dw}}(x) = 1 - 3.2x + 13.2x^2 - 20x^3 + 10x^4$ and the
non-symmetric profile $f_{\text{ns}}(x) = 1 + 0.5\,(e^{5x}-1)(1-x)$. The
computational results for the dispersion relation are given in
Fig.~\ref{fig:parameter_study}b-c. Furthermore, for $\kv$ in the
neighborhood of $\kv^\ast$ and $d/d_0=1.1$ we notice excellent agreement of
effective dispersion relation \eqref{eq:effective_dispersion_relation} with
the numerically computed curve $k_z(k_x)$ (Fig.~\ref{fig:enz_and_approx}).

%%%%%%%%%%%%%%%%%%%%%%%%%%%%%%%%%%%%%%%%%%%%%%%%%%%%%%%%%%%%%%%%%%%%%%%%%%%%%%%%
%%%%%%%%%%%%%%%%%%%%%%%%%%%%%%%%%%%%%%%%%%%%%%%%%%%%%%%%%%%%%%%%%%%%%%%%%%%%%%%%
%\paragraph{Comparison with direct numerical simulation.}

\begin{figure*}[tbph!]
\centering
\includegraphics[scale=0.33]{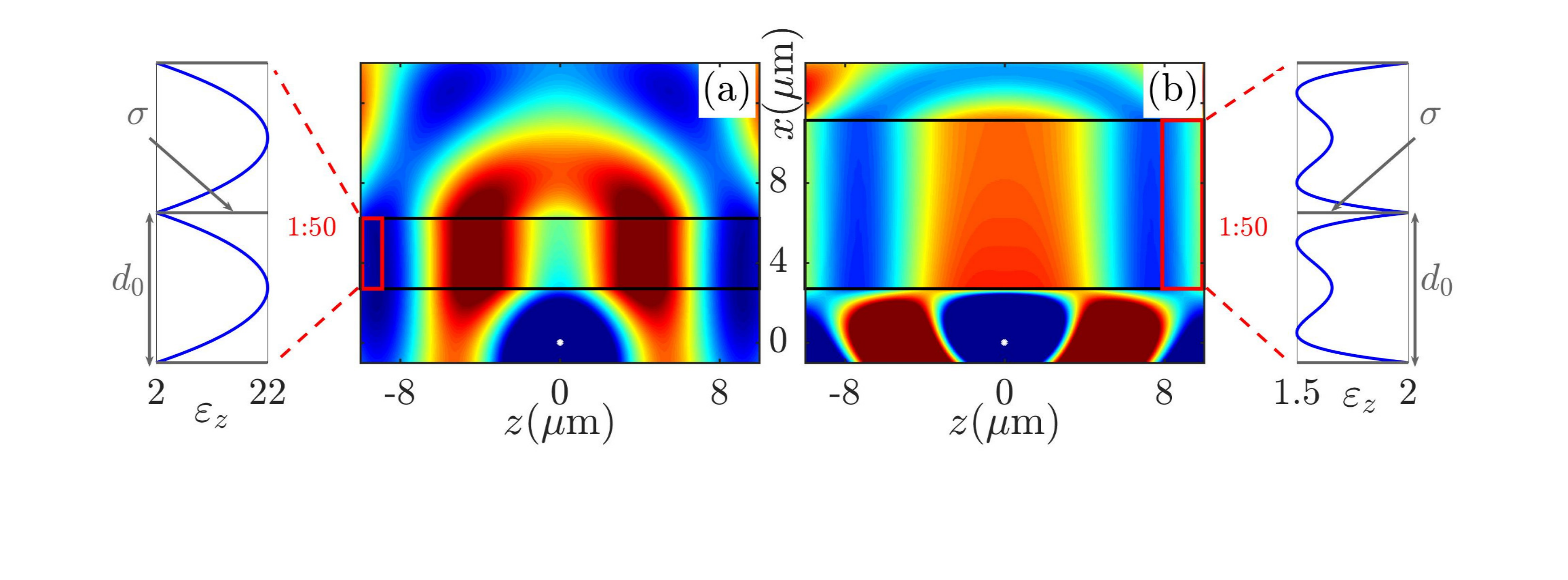}
\vspace{-2em}
\caption{Spatial distribution of $H_y$ in an anisotropic dielectric host
  with 100 layers of doped graphene with structural period
  $d=\text{Re}(d_{0})$ (black rectangle). A magnetic dipole source is
  located below the multilayer structures (white dots) emitting at $f=25$
  THz. The permittivity profile $\e_z(x)$ in (a) is a parabolic, and in (b)
  a double-well (insets). In the multilayer, waves propagate without
  dispersion and with no phase delay.}
\label{fig:simulations}
\end{figure*}
To test the results of our model against more practical
configurations, we carry out direct numerical simulations for a
system with a finite number of metallic
sheets. We choose graphene as the material for the 2D conducting
sheets, since it has been used extensively in plasmonic and optoelectronic
applications \cite{natPhot8_2014, natPhot6_2012}. In the THz frequency
regime, doped graphene behaves like a Drude metal because intraband
transitions are dominant \cite{natPhot6_2012, nature487_2012}. In this
frequency regime doped graphene supports plasmons \cite{natPhot6_2012}.
Hence, the conductivity of the metallic sheets is approximated by the Drude
formula, $\sigma = \im e^2 \mu_c/[\pi \hbar^2 (\omega+\im/\tau)]$. The
doping amount is $\mu_c=0.5$\,eV and the transport scattering time of
electrons is $\tau=0.5$\,ps to account for optical losses
\cite{mattheakis2016,wang2012}.

In Fig.~\ref{fig:simulations}, we present the spatial distribution of
$H_y(x,z)$ propagating through a structure of 100 graphene layers embedded
periodically in a lossless dielectric host with anisotropic and
spatial-dependent permittivity. The numerical computation is carried out for
parabolic profile \eqref{eq:parabolic_profile} with $\alpha=20/3$,
$\e_{z,0}=2$, $\e_x=1$, as well as the double-well profile, with
$\e_{z,0}=2$ and $\e_x=4$. By setting the structural period to $d=d_0$, we
observe the expected signature of ENZ behavior, namely, wave propagation
with no phase delay through the periodic structure \cite{prl97_2006,
natPhot9_2015,mattheakis2016}.

%%%%%%%%%%%%%%%%%%%%%%%%%%%%%%%%%%%%%%%%%%%%%%%%%%%%%%%%%%%%%%%%%%%%%%%%%%%%%%%%
%%%%%%%%%%%%%%%%%%%%%%%%%%%%%%%%%%%%%%%%%%%%%%%%%%%%%%%%%%%%%%%%%%%%%%%%%%%%%%%%
%%%%%%%%%%%%%%%%%%%%%%%%%%%%%%%%%%%%%%%%%%%%%%%%%%%%%%%%%%%%%%%%%%
%%%%%%%%%%%%%%%%%%%%%%%%%%%%%%%%%%%%%%%%%%%%%%%%%%%%%%%%%%%%%%%%%%
\section{Discussion and conclusion}
\label{sec:discussion}
%%%%%%%%%%%%%%%%%%%%%%%%%%%%%%%%%%%%%%%%%%%%%%%%%%%%%%%%%%%%%%%%%%
%%%%%%%%%%%%%%%%%%%%%%%%%%%%%%%%%%%%%%%%%%%%%%%%%%%%%%%%%%%%%%%%%%
In this section, we conclude our analysis by discussing implications of our
approach, summarizing our results and mentioning open related problems. Of
particular interest is a generalization of our result for the effective
dielectric permittivity of the layered plasmonic structure.

The notion of an effective permittivity $\e^{\text{eff}}_z$ that arises in
Eqs.~\eqref{eq:effective_dispersion_relation} and \eqref{eq:general_eeff}
bears a striking similarity to homogenization results for Maxwell's
equations~\cite{wellander2003}. In fact, it can be shown that
Eq.~\eqref{eq:general_eeff} can also be derived by applying an asymptotic
analysis procedure to the full system of time-harmonic Maxwell's equations.
For a general tensor-valued permittivity $\underline\e(x/d)$ and
sheet conductivity $\underline\sigma(x/d)$, the effective permittivity of
the metamaterial takes the form
\begin{align*}
  \underline\e^{\text{eff}} =
  \left<\underline\e\,\chi\right>_{\text{host}} +
  \frac{\im}{\omega}\left<\underline\sigma\,\chi\right>_{\text{sheet}}.
\end{align*}
Here, $\left<\,.\,\right>_R$ denotes the arithmetic average over region $R$
and $\chi$ is a weight function that solves a closed boundary value problem
in the individual layer \cite{maierxx}. In the special case of
$\underline\e=\text{diag}\,(\e_x,\e_z,\e_z),$ the weight function reduces
to the unit tensor, $\chi=I$. Understanding the ENZ behavior on the basis
of this more general effective permittivity is the subject of work in
progress.

Our work points to several open questions. For example, we analyzed wave
propagation through a plasmonic structure primarily {\em in absence of a
current-carrying source}. A related problem is to {\em analytically}
investigate how the dispersion band and ENZ condition derived here may
affect the modes excited by dipole sources located in the proximity of a
finite layered structure. This more demanding problem will be the subject
of future work.

%%%%%%%%%%%%%%%%%%%%%%%%%%%%%%%%%%%%%%%%%%%%%%%%%%%%%%%%%%%%%%%%%%%%%%%%%%%%%%%%
%%%%%%%%%%%%%%%%%%%%%%%%%%%%%%%%%%%%%%%%%%%%%%%%%%%%%%%%%%%%%%%%%%%%%%%%%%%%%%%%
%\paragraph{Conclusion.}
In conclusion, we have shown that dispersive Dirac cones are universal for
a wide class of plasmonic multilayer systems consisting of 2D metals with
isotropic, constant conductivity. We also derived a general, exact
condition on the structural period $d$ to obtain a corresponding dispersion
relation with ENZ behavior. The universality of our approach is key for the
investigation of wave coupling effects in discrete periodic systems and the
design of effective ENZ media. Our results pave the way to a systematic
study of homogenization and effective parameters in the context of more
general multilayer plasmonic systems.

%%%%%%%%%%%%%%%%%%%%%%%%%%%%%%%%%%%%%%%%%%%%%%%%%%%%%%%%%%%%%%%%%%%%%%%%%%%%%%%%
%%%%%%%%%%%%%%%%%%%%%%%%%%%%%%%%%%%%%%%%%%%%%%%%%%%%%%%%%%%%%%%%%%%%%%%%%%%%%%%%
\begin{acknowledgments}
We acknowledge support by ARO MURI Award No. W911NF-14-0247 (MMai, MMat,
EK, ML, DM); EFRI 2-DARE NSF Grant No. 1542807 (MMat); and NSF DMS-1412769
(DM). We used computational resources on the Odyssey cluster of the FAS
Research Computing Group at Harvard University.
\end{acknowledgments}

\medskip

%%%%%%%%%%%%%%%%%%%%%%%%%%%%%%%%%%%%%%%%%%%%%%%%%%%%%%%%%%%%%%%%%%%%%%%%%%%%%%%%
%%%%%%%%%%%%%%%%%%%%%%%%%%%%%%%%%%%%%%%%%%%%%%%%%%%%%%%%%%%%%%%%%%%%%%%%%%%%%%%%

\appendix

%%%%%%%%%%%%%%%%%%%%%%%%%%%%%%%%%%%%%%%%%%%%%%%%%%%%%%%%%%%%%
\section{Exact solution for parabolic dielectric profile}
\label{app:parabolic}
%%%%%%%%%%%%%%%%%%%%%%%%%%%%%%%%%%%%%%%%%%%%%%%%%%%%%%%%%%%%%

In this appendix, we outline the derivation of the exact dispersion
relation for parabolic dielectric profile~\eqref{eq:parabolic_profile}. As
a first step, we characterize the general solution of the differential
equation
\begin{align*}
  -\partial_x^2\E(x) + \kappa(k_z)\e_z(x) \E(x) = 0,
  \quad
  \kappa(k_z)=\frac{k_z^2-k_0^2\e_x}{\e_x},
\end{align*}
where the free space permittivity is set $\varepsilon_0=1$ and
$k_0=\omega\sqrt{\mu}$. In order to derive the solution of the above
differential equation, we apply a change of coordinate from $x$ to $\chi$,
viz.,
\begin{align*}
  x\to\chi=\rho\left(\frac xd-\frac12\right),
\end{align*}
using a complex-valued scaling parameter, $\rho$, to be determined below.
By identifying $\tilde\E(\chi)=\E(x)$ the differential equation now reads
\begin{align*}
  -\partial_\chi^2\tilde\E(\chi) +
  \kappa(k_z)\e_{z,0}\frac1{\rho^2} \Big(1 + \frac64\alpha -
  6\alpha\frac1{\rho^2}\chi^2\Big)\tilde\E(\chi) = 0.
\end{align*}
We now fix $\rho$ by the requirement that
\begin{align*}
  -6\alpha\frac1{\rho^4}{\kappa(k_z)}{\e_{z,0}}=\frac14.
\end{align*}
Thus, if $\alpha\leq0,$ we set
\begin{align}
\label{eq:rho}
  \rho(\alpha) = \Big(-24\,\alpha\,{\kappa(k_z)}{\e_{z,0}}\Big)^{1/4}.
\end{align}
We can analytically continue the above function $\rho(\alpha)$ to values
$\alpha>0$ by properly choosing one of the four branches of the (complex)
multiple-valued function $w(z)=z^{1/4}$. By the definition
\begin{align*}
  \nu = -1-\sqrt{{\kappa(k_z)}{\e_{z,0}}}\,
  \frac{1+(3/2)\alpha}{(-24\alpha)^{1/2}},
\end{align*}
the transformed differential equation for $\tilde \E(\chi)$ takes the
canonical form
\begin{align*}
  -\partial_\chi^2\tilde\E(\chi) + \Big(\frac12\chi^2-\nu-\frac12\Big)
  \tilde\E(\chi) = 0.
\end{align*}
This differential equation has the general solution
\begin{align}
  \tilde\E(\chi) = C_1 D_\nu(\chi) + C_2 D_\nu(-\chi),
  \label{eq:general_solution}
\end{align}
where $D_\nu(\chi)$ is the parabolic cylinder or Weber-Hermite function,
given by the formula
\begin{multline*}
  D_\nu(\chi) = 2^{\nu/2}e^{-\chi^2/4}
  \Big[\frac{\Gamma(1/2)}{\Gamma(1/2-\nu/2)}\Phi\big(-\nu/2,1/2;\chi^2/2\big)
    \\
    +\frac{\chi}{2^{1/2}}\frac{\Gamma(-1/2)}{\Gamma(-\nu/2)}
  \Phi\big(1/2-\nu/2,3/2;\chi^2/2\big)\Big],
\end{multline*}
and $C_1$ and $C_2$ are integration constants. In the above, $\Gamma(z)$ is
the Gamma function and $\Phi(a,b;z)$ is the confluent hypergeometric
function defined by the power series
\begin{align*}
  \Phi(a,b;z) =
  \sum_{n=0}^\infty \frac{(a)_n}{(b)_n}\frac{z^n}{n!},
\end{align*}
where $(a)_0=1$, $(a)_n=(a+n-1)(a)_{n-1}$ for $n\ge1$.

To derive the corresponding exact dispersion relation, we need to identify
the fundamental solutions $\tilde \E_{(j)}(x)$ ($j=1, 2$) and then
substitute general solution \eqref{eq:general_solution} written in terms of
these $\tilde \E_{(j)}(x)$ into determinant condition~(3). The resulting
condition reads
\begin{align*}
  D[\kv] &= \det
  \Bigg(
    e^{\im k_x d}
    \begin{bmatrix}
      1 & 0 \\
      - \im (\sigma/\omega)\kappa(k_z)& 1
    \end{bmatrix} \nonumber\\
   &\quad  -
    \begin{bmatrix}
      \tilde\E_{(1)}(\rho/2) & \tilde\E_{(2)}(\rho/2) \\
      \tilde\E'_{(1)}(\rho/2) & \tilde\E'_{(2)}(\rho/2)
    \end{bmatrix}
  \Bigg)
  \;=\; 0.
\end{align*}
After some algebra, the exact dispersion relation reads
\begin{multline}
  \cos(k_xd)+\frac{\Gamma(-\nu)}{\sqrt{2\pi}}
  \Big\{D_\nu(-\rho/2)D'_\nu(-\rho/2)\\
   + D_\nu(\rho/2)D'_\nu(\rho/2)
    \\
    - \frac{\kappa(k_z)\e_{z,0}\,\xi_0d}{2\rho}
    \big[D_\nu(\rho/2)^2-D_\nu(-\rho/2)^2\big]\Big\}=0.
  \label{eq:exact_dispersion}
\end{multline}
Here, $\xi_0=-\im\sigma/(\omega\e_{z,0})$ is the plasmonic thickness. Note
that, by our construction, $\rho$ and $\nu$ are $k_z$ dependent, viz.,
$\rho=\rho(k_z)$ and $\nu=\nu(k_z)$. Thus, Eq.~\eqref{eq:exact_dispersion}
still expresses an implicit relationship between $k_x$ and $k_z$. To
further simplify Eq.~\eqref{eq:exact_dispersion}, we expand $D_\nu(\rho/2)$
to fourth order in $z$. For sufficiently small structural period, $d$,
i.e., $|\kappa(k_z)d|\ll1$, and after some algebraic manipulations  the
exact dispersion relation simplifies to
\begin{align*}
  \cos(k_xd) & \approx 1 - \frac{\kappa(k_z)\e_{z,0}\xi_0d}{2} -
  (2\nu+1)(-(3/2)\alpha)^{1/2}\nonumber\\
  &\times \big(\sqrt{\kappa(k_z)\e_{z,0}}\,d\big)
  - \frac14\alpha\kappa(k_z)\e_{z,0}d^2.
\end{align*}
Furthermore, in the vicinity of Brillouin zone center, i.e., if $|k_x d|\ll
1$, we apply the Taylor expansion $\cos(k_xd)\approx 1 - (1/2)\,k_x^2d^2$
and use the definitions of $\nu$ and $\kappa(k_z)$ to obtain the {\em
effective} dispersion relation
\begin{align*}
  \frac{k_x^2}{\e_z^{\text{eff}}}
  \;+\;
  \frac{k_z^2}{\e_{x}}
  \;=\;
  k_0^2,
  \qquad
  \frac{\e_z^{\text{eff}}}{\e_{z,0}}
  =
  1+\alpha - \frac{\xi_0}{d},
\end{align*}
which is identical to Eq.~\eqref{eq:effective_dispersion_relation}.

%%%%%%%%%%%%%%%%%%%%%%%%%%%%%%%%%%%%%%%%%%%%%%%%%%%%%%%%%%%%%
\section{Numerical scheme for computation of dispersion bands}
\label{app:Newton_scheme}
%%%%%%%%%%%%%%%%%%%%%%%%%%%%%%%%%%%%%%%%%%%%%%%%%%%%%%%%%%%%%

In this appendix, we present more details on the numerical procedure to
compute dispersion bands for arbitrary dielectric profiles
$\varepsilon_z(x)$. For given problem parameters $\sigma$, $\omega,$ and
profile $\e_z(x)$, and fixed real $k_x$, consider the task of finding a
complex-valued solution $k_z$ of \eqref{eq:determinant_condition}.

We formulate a Newton method in order to solve the implicit dispersion
relation $D[\kv]=0$ numerically. For this purpose, we first need to
characterize the variation $\delta\E_{(i)}$ of solutions $\E_{(i)}$ of
Eq.~\eqref{eq:governing_ode} with respect to $k_z$. We make the observation
that $\delta\E_{(i)}$ is the unique solution of the differential equation
\begin{align*}
  -\partial_x^2\delta\E_{(i)} + \kappa(k_z)\e_z(x) \delta\E_{(i)} +
  \kappa'(k_z)\e_z(x) \E_{(i)} = 0,
  \end{align*}
  where
  \begin{align*}
 \kappa'(k_z)=\frac{2k_z}{\e_x},
  \\
  \delta\E_{(i)}(0) = 0, \; \delta\E'_{(i)}(0) &= 0.
\end{align*}
With this prerequisite at hand, the variation of $D[\kv]$ with respect to
$k_z$ can be expressed as follows:
\begin{multline}
  \delta D[\kv] =
  -e^{\im k_xd}\Big\{
  \delta\E_{(1)}(d)\,\big(1-\E'_{(2)}(d)\big)\\
  + \big(1-\E_{(1)}(d)\big)\,\delta\E'_{(2)}(d)
  \\
  + \big(\im(\sigma/\omega)\kappa'(k_z) + \delta\E'_{(1)}(d)\big)\,\E_{(2)}(d)
  \\
  + \big(\im(\sigma/\omega)\kappa(k_z)+\E'_{(1)}(d)\big)\delta\E_{(2)}(d)
  \Big\}.
  \label{eq:variation_D}
\end{multline}

Next, we outline the steps of the Newton scheme. Let $k_x$ be fixed.
Suppose that starting from an initial guess $k_z^{(0)}$ we have computed an
approximate solution $k_z^{(n)}$ of Eq.~\eqref{eq:determinant_condition}.
We then compute a new approximation $k_z^{(n+1)}$ according to the
following sequence of steps:
\medskip

\begin{itemize}
  \item Solve the first order systems ($i=1,2$)
    \begin{align*}
      &\begin{cases}
        \begin{aligned}
          -\partial_x(\E'_{(i)}) + \kappa(k_z)\e_z(x) \E_{(i)}
          &= 0,\\
          -\partial_x(\E_{(i)}) &= \E'_{(i)},
        \end{aligned}
      \end{cases}
    \end{align*}
    with initial conditions $\E_{(1)}(0) = 1$, $\E'_{(1)}(0) = 0$,
    $\E_{(2)}(0) = 0$, $\E'_{(2)}(0) = 1$.
  \item Solve the systems ($i=1,2$)
    \begin{align*}
      &\begin{cases}
        \begin{aligned}
          -\partial_x(\delta\E'_{(i)}) + \kappa(k_z)\e_z(x) \delta\E_{(i)} +
          \kappa'(k_z)\e_z(x) \E_{(i)} &= 0,
          \\
          -\partial_x(\delta\E_{(i)}) &= \delta\E'_{(i)},
          \\
          \partial\E_{(i)}(0) = 0, \; \partial\E'_{(i)}(0) &= 0.
        \end{aligned}
      \end{cases}
    \end{align*}
  \item
    Compute $D[k_x,k_z^{(n)}]$ and $\delta D[k_x,k_z^{(n)}]$ given by
    Eqs.~\eqref{eq:determinant_condition} and \eqref{eq:variation_D}.
  \item
    Update:
    \begin{align*}
       k_z^{(n+1)} = k_z^{(n)} - \frac{D[k_x,k_z^{(n)}]}{\delta D[k_x,k_z^{(n)}]}.
    \end{align*}
\end{itemize}

%%%%%%%%%%%%%%%%%%%%%%%%%%%%%%%%%%%%%%%%%%%%%%%%%%%%%%%%%%%%%%%%%%%%%%%%%%%%%%%%
%%%%%%%%%%%%%%%%%%%%%%%%%%%%%%%%%%%%%%%%%%%%%%%%%%%%%%%%%%%%%%%%%%%%%%%%%%%%%%%%

%merlin.mbs apsrev4-1.bst 2010-07-25 4.21a (PWD, AO, DPC) hacked
%Control: key (0)
%Control: author (8) initials jnrlst
%Control: editor formatted (1) identically to author
%Control: production of article title (-1) disabled
%Control: page (0) single
%Control: year (1) truncated
%Control: production of eprint (0) enabled
%

\end{document}